\documentclass[a4]{article}
\usepackage[left=2cm,right=2cm,top=2cm,bottom=2cm,bindingoffset=0cm]{geometry}
\usepackage{epsfig}
\usepackage{epstopdf}
\usepackage{multirow}
\newcommand{\Ref}[1]{(\ref{#1})}
\newcommand{\beao}{\begin{eqnarray*}}
\newcommand{\eeao}{\end{eqnarray*}}
\newcommand{\be}{\begin{equation}}
\newcommand{\ee}{\end{equation}}
\newcommand{\bea}{\begin{eqnarray}}
\newcommand{\eea}{\end{eqnarray}}
\newcommand{\beq}{\begin{eqnarray}}
\newcommand{\eeq}{\end{eqnarray}}
\newcommand{\nn}{\nonumber}

\newcommand{\la}{\lambda}

\begin{document}

\title{{Induced color charges,
effective $\gamma \gamma g$- vertex in QGP.\\ Applications to
heavy-ion collisions}}
\author{
  V. Skalozub \thanks{e-mail:Skalozubv@daad-alumni.de}\\
{\small Oles Honchar Dnipro National University, 49010 Dnipro, Ukraine}}

\date{}
\maketitle.

%

%
\begin{abstract}
We calculate the induced color charges $Q_{ind.}^3, Q_{ind.}^8 $
and the effective vertex $\gamma - \gamma -  $gluon  generated in
quark-gluon plasma with the $A_0$ condensate because of the color
parity violation at this background.  To imitate  the case of
heavy-ion collisions, we consider the model of the plasma confined
in the narrow infinite plate and derive the classical gluon
potentials $\bar{\phi}^3$, $\bar{\phi}^8$ produced by these
charges. Two applications - scattering of photons on the plasma
and conversion of gluon fields in two photons radiated from the
plasma - are discussed.

\end{abstract}
\section{Introduction}
Investigations of the deconfinement phase transition (DPT) and
quark-gluon plasma (QGP)   are in the center of modern high energy
physics. These phenomena  happen at high temperature due to
asymptotic freedom of strong interactions. The researches are
carried out either in experiments on hadron collisions  or in
quantum field theory. The order parameter for the DPT is
Polyakov's loop (PL), which is zero at low temperature and nonzero
at high temperatures $T > T_d $, where $T_d \sim 160-180 $ MeV
\cite{Fodor2009} is the phase transition temperature. Standard
information on the DPT is adduced, in particular, in
\cite{Gross1981}, \cite{Satz}, \cite{Greensite}.

    The PL is defined as \cite{S}:
\be \label{PL} PL = \int_C d x_4 ~A_0(x_4, \vec{x}) . \ee
Here $A_0(x_4, \vec{x})$ is the zero component of the gauge field
potential, the integration contour is going along the fourth
direction  and back to an initial point in the lattice Euclidean
space-time. The PL was introduced in pure gluodynamics. It
violates the center of the color group symmetry $Z(3)$ that
results in non conservation of the color charges $Q^3, Q^8 $.

The QGP state  consists of  quarks and gluons liberated from
hadrons.   Polyakov's loop is not a solution to local Yang-Mills
equations. Instead the local order parameter for the DPT is the
$A_0$ condensate, which is a constant   at $T > T_d$.  It can be
calculated, in particular, from a two-loop effective potential.
First this has been done by Anishetty \cite{Anishetty1984}. More
details on different type     calculations carried out in analytic
quantum field theory see in \cite{Skalozub1995}. In particular,
important point is a gauge-fixing invariance, which was proved via
Nielsen's identity approach.   Taking into consideration these
results, we have to consider  QGP as  the state  at the $A_0$
background which breaks color symmetry. So, new type phenomena may
happen.

Already, in Refs.\cite{Kalashnikov94}, \cite{Kalashnikov96}  in
pure SU(2) gluodynamics   the gluon spectra at the $A_0$ have been
calculated and investigated.  In particular, the  induced color
charge $Q^3_{ind.} $ was also  computed. It  was shown that the
state with the condensate is free of infrared instabilities
existed in gluon plasma in empty space. So, any kind resummations
of infrared singularities are not needed.  It has been
demonstrated that the consistent Slavnov-Taylor identity must
include the induced charge and corresponding current. Thus, the
ground state with the $A_0$ is a good approximation to the plasma
after $DPT$.

In Ref. \cite{Baranov18} the induced charges $Q^3_{ind.},
Q^8_{ind.} $ generated by quark loops in $QCD$ have also been
calculated. They are additive quantities  determined as the  sum
over all the gluon and fermion species.  In what follows, we
consider the $QCD$ case but the precise  values of the induced
charges will not be specified. That is, we concentrate on the
qualitative picture of the phenomena investigated.
 To be self-contained, we begin with a  brief calculation of the
induced charge generated in the quark sector. Then we investigate
a number of effects happened due to this charge. The
imaginary-time formalism at the $A_0$ = const background is used.

In Sect. 2 the color induced charges $Q_{ind.}^3, Q_{ind.}^8 $,
generated by tadpole quark loops with one gluon lines, which are
nonzero because of Furry's theorem violation, are calculated. Here
we also develop the procedure of accounting for the induced
charges via Schwinger-Dyson's equation. In Sect. 3 we consider a
simple model of the plasma confined in a nero in one dimension and
infinite in two other dimensions plate with the $A_0$ condensate
and induced charges. It imitates the case of $QGP$ created in
heavy-ion collisions. We compute the classical gluon potentials
$\bar{\phi}^3$ and $\bar{\phi}^8$ generated by the induced charges
$Q_{ind.}^3, Q_{ind.}^8 $. In Sect. 4 the effective $\gamma \gamma
 G$ vertex generated in the plasma is calculated in high temperature
approximation.   As  applications of the potentials $\bar{\phi}^3$
and $\bar{\phi}^8$, in Sect. 5  the processes of photon scattering
on these potentials and conversion of gluons in two photons are
considered. These new phenomena have to happen due to the
three-linear   effective vertexes. The necessary notations and
definitions are given in Appendix.
\section{Induced color charges and quark propagator}
In what follows, we consider the case of $A_0^3$ background field
and present the color field in the form $Q^a_\mu \to A_0 \delta^{a
3}  \delta_{\mu 4} + Q^a_\mu $, where $Q^a_\mu $ is a quantum
field. The Eqs.\Ref{Ge}, \Ref{vertQe} taken in momentum space will
be used to calculate the induced color charge $Q_{ind.}^3$. The
calculation of $Q_{ind.}^8$ is similar (see \cite{Baranov18}) and
the final results will be adduced, only.

The explicit expression for these definitions  is given by the
expression $Q^a_\mu Q_{ind.}^3 \delta_{\mu 4}\delta_{a 3}$ =
$Q^3_4 Q_{ind.}^3$, where
\be \label{Q1}Q_{ind.}^3 = \frac{g}{\beta}\sum_{p_4} \int
\frac{d^3p}{(2 \pi)^3}Tr [\gamma^4
\frac{\lambda^3_{ij}}{2}G^{ij}(p_4, \vec{p}, A_0)].\ee
Here, the sign "-" was accounted for because of the fermion loop.
$\lambda^3$ is Gell-Mann matrix, $\beta = 1/T $ is inverse
temperature. The expressions for the propagators are
\be \label{FG} G^{11}= \frac{\gamma^4 (p_4 - A_0) + \vec{p}\cdot
\vec{\gamma} + m}{(p_4 - A_0)^2 + \vec{p}^2 + m^2}, ~G^{22}=
\frac{\gamma^4 (p_4 + A_0) + \vec{p}\cdot \vec{\gamma} + m}{(p_4 +
A_0)^2 + \vec{p}^2 + m^2}. \ee
Here, for brevity, we denoted as $A_0 = \frac{g A_0}{2}$ entering
the interaction Lagrangian.
Accounting for the trace  $Tr [(\gamma^4)^2] = - 4$, diagonal
values of $\lambda^3$ and $Tr [\gamma^4 \vec{\gamma} ] = 0$ we get
\be \label{Q2}Q_{ind.}^3 = - \frac{2 g}{\beta}\sum_{p_4} \int
\frac{d^3p}{(2 \pi)^3}[\frac{(p_4 - A_0) }{(p_4 - A_0)^2 +
\vec{p}^2 + m^2}- \frac{(p_4 + A_0) }{(p_4 + A_0)^2 + \vec{p}^2 +
m^2}].\ee
To unite these two terms we make the change $p_4 \to - p_4$ in the
first term and obtain
\be \label{Q3}Q_{ind.}^3 =  \frac{4 g}{\beta}\sum_{p_4} \int
\frac{d^3p}{(2 \pi)^3}  \frac{p_4 + A_0 }{(p_4 + A_0)^2 +
\vec{p}^2 + m^2}.\ee
The sum over $p_4 = \frac{\pi (2 n + 1)}{2 \beta }$ can be
calculated by using formula
\be \label{sum} \frac{1}{\beta}\sum_{p_4}f(p_4) = - \frac{1}{4 \pi
i} \int_C \tan [\frac{\beta \omega}{2}] f(\omega), \ee
where contour C encloses clockwise the real axis in the complex
plane $\omega$.

The straightforward calculations (after transformation to the
spherical coordinates and angular integrations) give
\be \label{Q4}Q_{ind.}^3 = \frac{g \sin (A_0 \beta)}{\pi^2}
\int_0^{\infty} p^2 d p \frac{1}{\cos (A_0 \beta)+ \cosh
(\epsilon_p \beta)},\ee
where $\epsilon_p^2 = p^2 + m^2$.

Considering the limit of high temperature $\beta \to \infty$ we
get
\be \label{Q3T} Q^3_{ind.} = g A_0 \bigl[ \frac{4}{3} \beta^{-2} -
\frac{2 m^2}{3 \pi^2} \beta + O (\beta^3) \bigr ].\ee
Hence we see that the first term is independent of the mass and
dominant at high temperature.

Now, for completeness,  we  calculate  the temperature sum in
Eq.\Ref{Q3}.

The integrand in Eq.\Ref{Q3} has the form
\be \label{f4} f(p_4) =  \frac{p_4 + A_0 }{(p_4 - p_4^{(1)})(p_4 -
p_4^{(2)}) },\ee
where $p_4^{(1)} = - A_0 + i \epsilon_p, p_4^{(2)} = - A_0 - i
\epsilon_p$. The sum in Eq.\Ref{sum} after computing the simple
residues  equals to
\be \label{sum1} S_1 = \frac{1}{\beta}\sum_{p_4}f(p_4) = -
\frac{1}{2} [\frac{i \epsilon_p}{p_4^{(1)} - p_4^{(2)}}
\tan(\frac{\beta}{2}p_4^{(1)}) + \frac{- i \epsilon_p}{p_4^{(2)} -
p_4^{(1)}} \tan(\frac{\beta}{2}p_4^{(2)})]. \ee
Substituting the corresponding parameters and fulfilling
elementary transformations we find
\be \label{sum2} S_1 = \frac{1}{2} \frac{ \sin(A_0 \beta)}{\cos
(A_0 \beta)+ \cosh(\epsilon_p \beta)}. \ee
By substituting $S_1$  in Eq.\Ref{Q3} we obtain the result
Eq.\Ref{Q4}.

Performing similar calculations for the $Q^8_{ind.}$ we obtain
\cite{Baranov18}
\be \label{Q8T} Q^8_{ind.} = g A_0^8 \bigl[ \frac{16}{3\sqrt{3}}
\beta^{-2} - \frac{8 m^2}{3\sqrt{3}\pi^2} \beta + O (\beta^3)
\bigr ].\ee
Here, $A_0^8$ is background field generated in the plasma. For our
problem, it is a given number.

Now, we calculate the quark propagator accounting for the induced
charge by means of Schwinger-Dyson's equation. In the Euclid
space-time it reads
\be \label{SDe} S^{-1}(p) = - ( \gamma^4 ( p_4 -
\frac{\lambda^3}{2} g A_0) + \vec{\gamma}\vec{p} ) + m - \Sigma
(p), \ee
where $\Sigma (p)$ is a quark mass operator.  In our problem, to
take into consideration the presence of the induced charge, we
separate the part of radiation corrections $\Sigma^{(tp.)} $
equaling to the sum of the tadpole diagrams
 with one gluon line $G^3_4$, which relates the quark
bubble with a quark line. In Eq.\Ref{SDe} we also substitute the
$A_0$ expression explicitly. In the rest frame of the plasma,
where actual calculations are carried out, the velocity vector is
$u_\mu = (u_4 = 1, \vec{u} = 0) $.

Next we have to take into consideration  the gluon field
propagator $G^3_{4 4}(k)$. For that we use the generalized Green's
function  of neutral gluons. It reads (in the Lorentz-Feynman
gauge) \cite{Kalashnikov94}, \cite{Kalashnikov96}
\be \label{G44} (G^3_{4 4})^{-1} = k^2 - \Pi_{4 4 }(k_4,
\vec{k}),\ee
where $\Pi_{4 4 }(k^2) $ is $4-4 $ component of the polarization
tensor. For $k_4 = 0,\vec{k} \to 0$ it defines Debye's temperature
mass having the order  $m^2_D \sim g^2 T^2$. This mass is
responsible for screening of the Coulomb color fields.

 The  component  of interest $G_{4
4}^3 $ taken at zero momenta reads \cite{Kalashnikov94},
\cite{Kalashnikov96}
\be \label{G3}G_{4 4}^3(p = 0) = \frac{1}{m^2_D}. \ee
%
Taking into consideration Eqs. \Ref{vertQe}, \Ref{Q4}, \Ref{G3},
we obtain
\be \label{Sigma} \Sigma^{(tp.)} = - \frac{\lambda^3}{2} \gamma^4
\frac{g Q_{ind.}^3 }{m^2_D}. \ee
Substituting this in Eq.\Ref{SDe} we conclude that the resummation
of tadpole insertions results in the replacement of $g A_0
 \to g A_0  + g \frac{Q_{ind.}^3}{m^2_D} $ in the initial propagator.
Since $ Q_{ind.}^3$ is positive, the account of the induced charge
in quark propagation looks as increasing the value of the $A_0$
condensate. Similar picture takes place for the $A_0^8$ background
field.  There are no needs in including these graphs in actual
calculations. That can be done in final expressions obtained from
other type diagrams.
\section{Potentials of classical color fields}
The presence of the induced color charges in the plasma leads to
the generation of classical gluon potentials. To describe
 this phenomenon we introduce  a simple
model motivated by heavy-ion collisions. In this case the plasma
is created for a short period of time in a finite space volume
which has a much smaller size in the direction of collisions
 compared to the transversal  ones.

 We consider  the QGP confined in the narrow plate
of the size L in z-axis direction and infinite in x-, y-
directions. For this geometry, we calculate the classical
potentials $\bar{\phi}^3 = G^3_4, \bar{\phi}^8 = G^8_4$ by solving
the classical field equations for the gluon fields $G^3_4, G^8_4$
generated by the corresponding induced charges $Q_{ind.}^3,
Q_{ind.}^8$. In doing so we take into consideration the results of
Refs. \cite{Kalashnikov94}, \cite{Kalashnikov96} where the gluon
modes at the $A_0$ background have been calculated. Either
transversal or longitudinal modes were derived. For our problem,
we are interested in the latter ones. The longitudinal modes of
fields $G^3_4, G^8_4$ have temperature masses $~\sim g^2 T^2$ They
are not affected by the background fields.

Therefore, the classical potential $\bar{\phi}^3$  is calculated
from the equation
\be \label{Pois3} [\frac{\partial^2}{\partial x_\mu^2} - m^2_D]
\bar{\phi}^3 = - Q_{ind.}^3. \ee
And similar equation for $\bar{\phi}^8$.

Making Fourier's transformation to momentum $k$-space we derive
the spectrum of modes - $k^2_4 = k_x^2 + k_y^2 + k^2_z + m^2_D$,
where $k^2_z = (\frac{2 \pi}{L})^2 l^2$ and $l = 0, \pm 1, \pm
2,...$. The discreteness of $k_z$ is due to the periodic boundary
condition for the plane: $\bar{\phi}^3(z) = \bar{\phi}^3(z + L)$.
The general solution to  Eq.\Ref{Pois3} is
\be \label{Gsolush} \bar{\phi}^3(x_4, \vec{x}) = d + a~ e^{-i (k_4
x_4 - \vec{k}\cdot \vec{x})} + b~ e^{i (k_4 x_4 - \vec{k}\cdot
\vec{x})}. \ee
In case of zero induced charge d = 0, and we have two well known
 plasmon modes. In case of $ Q_{ind.}^3 \not = 0$,
 the values  $a, b, d$  calculated
from the confinement boundary condition
\be \label{confcond} \bar{\phi}^3(z = - \frac{L}{2}) =
\bar{\phi}^3(z =  \frac{L}{2}) \ee
result in the expression
\be \label{solution} \bar{\phi}^3(z) \frac{Q_{ind.}^3}{m^2_D}
\bigl[ 1 - \frac{\cos (k_z z)}{\cos( k_z L/2)} \bigr ]. \ee
The generated potential depends on the z-variable only.  There are
no dynamical plasmon states at all. The same result follows for
the potential $\bar{\phi}^8(z) $. This is the main observation. In
the presence of the induced charges the static  classical color
potentials have to present in the  plasma.

For  applications it is also necessary to get Fourier's transform
$\bar{\phi}^3(k)$ of the potential \Ref{solution} to momentum
space k. Fulfilling that for the interval of z $[- \frac{L}{2},
\frac{L}{2}]$ we obtain
\be \label{Fphi3}\bar{\phi}^3(k) = \frac{Q_{ind.}^3 L}{m^2_D}~~
\frac{\sin(k L/2)}{(k L/2)} \frac{k^2_z}{k^2_z - k^2}, \ee
where the values of $k_z $ are adduced after Eq.\Ref{Pois3}.

The energy for a  one mode with momentum $k_z$ is positive and
equals to
\be \label{emode} E_l =
\frac{(Q_{ind.}^3)^2}{m^4_D}\frac{k^2_z}{2} L =
\frac{(Q_{ind.}^3)^2}{m^4_D} \frac{2 \pi^2}{L} l^2. \ee
The total energy is given by the sum over $l$ of energies
\Ref{emode}. Similar results hold for the potential
$\bar{\phi}^8$.

As  we derived, in the presence of the induced charges the static
gluon potentials with positive energy should be generated.
Dynamical longitudinal modes do not exist. This is the consequence
of the  condition Eq.\Ref{confcond}. Obvious that such a situation
is independent of the bag form where the plasma is confined. In
general, we have to expect that  the color static potentials
$\bar{\phi}^3$, $\bar{\phi}^8$ have to exist in the $QGP$ and
produce specific processes. Some of them we consider below.
\section{Effective $\gamma \gamma G $  vertexes in $QGP$}
Other interesting objects which have to be generated in the $QGP$
with $A_0$ condensate are the effective three-line vertexes
$\gamma \gamma G^3, \gamma \gamma G^8. $ They  should exist
because of Furry's theorem violation and relate colored and white
states. These   vertexes, in particular,  have to result in
observable processes of new type - inelastic scattering of
photons, splitting (dissociation or conversion) of gluon
$\bar{\phi}^3$, $\bar{\phi}^8$ potentials in two photons.

In this and next sections we calculate the vertex $\gamma \gamma
G^3$ and investigate these processes in the plasma.


\begin{figure}[h!]
\centering
\includegraphics{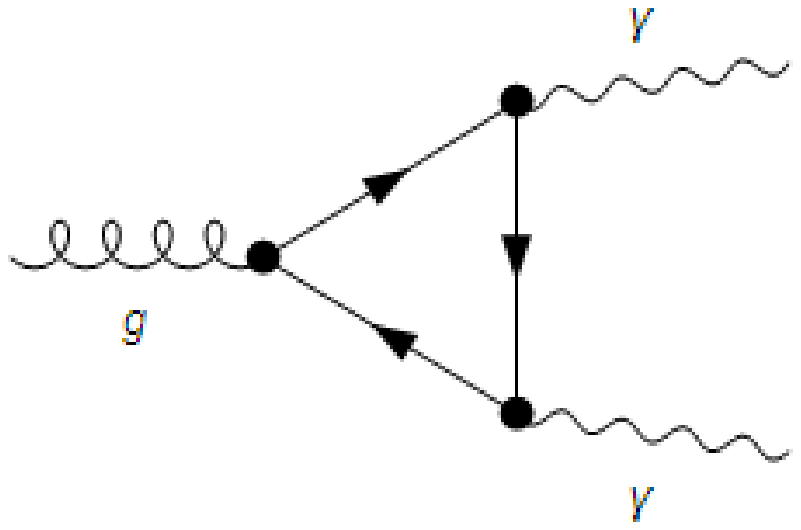}
\end{figure}
%

 Let us consider
the   vertex $\Gamma_{\mu \la}^\nu$ depicted in the plot. It
consists of two diagrams. The second diagram is obtained by
changing the direction of the quark line.  We use the notations:
all the momenta are ingoing, first photon $\gamma_1 (k^1_\mu)$,
second photon $\gamma_2 (k^3_\la)$, color a=3 gluon $Q^3 (k^2_\nu)
$, and $k^1 + k^2 + k^3 = 0$. $k^{1,2,3}$ are momenta of external
fields.

We consider the contributions coming from the traces of four
$\gamma$-matrixes, which are   proportional to the quark mass and
dominant for small photon momenta $k^1, k^3 << m$. The
corresponding expression reads
\be \label{gamma}  \Gamma_{\mu \la}^\nu (k^1, k^3) =  \Gamma_{\mu
\la}^{\nu, (1)}(k^1, k^3) +   \Gamma_{\mu \la}^{\nu, (2)}(k^1,
k^3), \ee
where
\be \label{gamma1}  \Gamma_{\mu \la}^{\nu, (1)}(k^1, k^3) =
\frac{1}{\beta} \sum_{p_4} \int \frac{d^3 p}{(2 \pi)^3}
\frac{N_{1}}{D(\tilde{P}) D(\tilde{P}- k^1) D(\tilde{P} +
k^3)}.\ee
Here $\beta = T^{-1}$ is inverse temperature, summation is over
$p_4 = 2\pi T ( l + 1/2), l = 0, \pm 1, \pm 2, ...$,  integration
is over three dimensional momentum space $p$, $N_1$ denotes the
numerator coming from   the first diagram, $\tilde{P} =
(\tilde{P}_4 = p_4 - A_0 , \vec{p})$, $D(\tilde{P}) = ( p_4 -
A_0)^2 + \vec{p}^2 + m^2  = \tilde{P}_4 ^2 +  \epsilon_p^2 $ and $
\epsilon_p^2 = \vec{p}^2 + m^2$ is energy of free quark squared.
The functions $ D(\tilde{P}- k^1),  D(\tilde{P} + k^3) $ assume a
corresponding shift in momentum. The numerator $N_1$ is
\be \label{N1}( N_1 )_{\mu\nu\la} = \delta_{\mu\nu} (\tilde{P} -
k^2)_\la + \delta_{\la\nu} (\tilde{P} - k^2)_\mu + \delta_{\mu\la}
(\tilde{P} - q)_\nu, \ee
where $ q = k^3 - k^1$ is photon momentum   transferred.

The expression for the second term in \Ref{gamma} is coming from
the second diagram and obtained from  \Ref{gamma1}, \Ref{N1} by
substitutions $k^1 \to - k^1, k^2 \to - k^2, q \to - q$. We denote
the second numerator as $N_2$. In what follows we carry out actual
calculations for the first term in \Ref{gamma} and adduce the
results for the second one.

Now, we  take into consideration the fact that in the high
temperature limit the large  values of integration  momentum $p$
give the leading contribution. Therefore we can present the
functions $D(\tilde{P}),  D(\tilde{P}- k^1),  D(\tilde{P} + k^3)$
in the form:
\bea \label{Di}  D(\tilde{P})  & =& \tilde{P}_4 ^2 +  \epsilon_p^2
=  \tilde{P} ^2, \\ \nn
 D(\tilde{P}- k^1) &=&  \tilde{P} ^2 \Bigl(1-  \frac{2  \tilde{P}\cdot k^1 - k_1^2}{\tilde{P} ^2} \Bigr), \\ \nn
   D(\tilde{P}+ k^3) &=&  \tilde{P} ^2 \Bigl( 1+  \frac{2  \tilde{P}\cdot k^3 + k_3^2}{\tilde{P} ^2} \Bigr). \eea
Here, $k_1^2 = (k^1_4)^2 + \vec{k}_1^2,   k_3^2 = (k^3_4)^2 +
\vec{k}_3^2$. At high  temperature and $ \tilde{P} ^2  \to \infty$
the k-dependent terms are  small.   So, we can expand in these
parameters and obtain for the integrand  Intd.  in Eq.\Ref{gamma1}
%
%
\be \label{Intd} Intd.  = \frac{N_1}{(\tilde{P} ^2)^3} \Bigl[ 1 +
\sum_{i=1}^4 A_i \Bigr], \ee
where \be \label{Ai} A_1 = - 2 \frac{ (\tilde{P}\cdot
q)}{\tilde{P}^2}, ~~A_2 =  - \frac{k^2_3 - k_1^2}{\tilde{P}^2} \ee
\be \label{Aj}  A_3 = - 4 \frac{(\tilde{P}\cdot k^1)
(\tilde{P}\cdot k^3)}{\tilde{P} ^2}, ~~ A_4 =  4
\frac{(\tilde{P}\cdot k^1)^2 + (\tilde{P}\cdot k^3)^2}{\tilde{P}
^2} \ee \nn
and vector $q_\mu = (q_4, \vec{q})$.

For the  second diagram we have to substitute $q \to - q$,  other
terms are even and do not change.

Further we concentrate on the scattering of photons on the
potential $Q_4^3$ in the  medium rest frame and set the thermostat
velocity $u_\nu = (1, \vec{0}), \nu = 4$. The corresponding terms
in the numerators are
\be \label{Ns} N_1 -> \delta_{\mu\la} (\tilde{P} + q )_4, ~~  N_2
-> \delta_{\mu\la} (\tilde{P} - q )_4, \ee
remind that $\tilde{P}_4 = p_4 - A_0$  and $\tilde{P}^2 = ( p_4 -
A_0)^2 + \epsilon_p^2$.

We have to calculate in general the series of two types
corresponding to these numerators:
\be \label{Si} S_1^{(n)} =\frac{1}{\beta} \sum_{p_4} \frac{p_4 -
A_0}{(\tilde{P}^2 )^n},  ~~ S_2^{(n)} =\frac{1}{\beta} \sum_{p_4}
\frac{q_4}{(\tilde{P}^2 )^n}, n = 3, 4, 5. \ee
These functions can be calculated from the $ S_1^{(1)} $ and $
S_2^{(1)}$ by computing  a number of derivatives over
$\epsilon_p^2$. The latter series result in simple expressions.
First is the one calculated already  for the tadpole diagram
Eq.\Ref{sum2}. But now we have to change the sing $A_0 \to - A_0$.
%
%
The function $  S_2^{(1)}$ is
\be \label{21} S_2^{(1)} = \frac{1}{\beta} \sum_{p_4}
\frac{q_4}{\tilde{P}^2 } = - \frac{1}{2 \epsilon_p}
\frac{\sinh(\epsilon_p \beta)}{\cos (A_0 \beta) + \cosh(\epsilon_p
\beta)}. \ee
Hence, the explicit analytic expressions can be obtained for the
integrand and integrations  over $d^3 p$ carry out in terms of
known  functions and their derivatives.


%
%

Let us adduce the  expressions for $A_i$ obtained after some
simplifying algebraic transformations:
 \be \label{A1} A_1 = - 2
\frac{(p_4 - A_0) q_4}{\tilde{P} ^2},\ee
\be \label{A2} A_3 = -\frac{ 4}{\tilde{P} ^2} \Bigl[ (1 -
\frac{\epsilon^2_p}{  \tilde{P} ^2}) k^1_4 k^3_4 +
\frac{(\vec{p}\cdot \vec{k}_1)  ( \vec{p}\cdot \vec{k}_3)}{
\tilde{P} ^2} \Bigr],\ee
\be \label{A4} A_4 =\frac{ 4}{\tilde{P} ^2} \Bigl[ (1 -
\frac{\epsilon^2_p}{  \tilde{P} ^2})(( k^1_4)^2+ ( k^3_4)^2) +
\frac{(\vec{p}\cdot \vec{k}_1)^2 +  ( \vec{p}\cdot \vec{k}_3)^2 }{
\tilde{P} ^2} \Bigr],\ee
Accounting for the structure of the numerators \Ref{Ns}, we see
that  the terms without $\tilde{P}_4$ are canceled in the sum of
two diagrams and the resulting amplitude consists of the terms
\be \label{M1} M_1 = 2 \delta_{\mu\la} \frac{p_4 - A_0}{(\tilde{P}
^2)^3} ( 1 + A_1 + A_3 + A_4  ) \ee
and \be \label{M2} M_2 = - 4 \delta_{\mu\la} \frac{(p_4 - A_0)
q^2_4}{(\tilde{P} ^2)^4}. \ee
Thus, all the contributions of the $S_2^{(n)}$ series are canceled
in the total. Now we tern to $d^3 p$ integration.
%
%

The expressions in Eqs.\Ref{M1},\Ref{M2} contain different powers
of $\tilde{P} ^2$, and hence different corresponding powers of
$\beta$ appear even in the leading $p \to \infty$ approximation.
This corresponds to the first term in the expansion $\epsilon_p =
p + \frac{1}{2} \frac{m^2}{p} + O(p^{-3})$. Below, we carry out
integration in this leading in $T \to \infty$ approximation.

We present our procedure considering the first term in Eq.\Ref{M1}
which corresponds to the  second derivative of $S_1^{(1)}$ over
$\epsilon_p^2$ and  equals  to
\be \label{S3} S_3 = - A_0 \beta ~\frac{Sech(\beta \epsilon_p
/2)^4}{64 p^3} ( - 2 \beta \epsilon_p + \beta \epsilon_p
Cosh(\beta \epsilon_p ) +  Sinh(\beta \epsilon_p ) ). \ee
Then in the spherical coordinates we calculate the integral
\be \label{I3} I_3 = \int\limits_{-\infty}^{\infty} d^3 p~ S_3  =
4 \pi   \int\limits_{0}^{\infty  }  p^2 d p  ~S_3(p). \ee
 In leading approximation $\epsilon_p \beta = p \beta$.   Making
the change of variables $p \beta = y$ we obtain for  \Ref{I3},
\be \label{I3a} I_3 = - \frac{A_0 \pi \beta}{16}
\int\limits_{0}^{\infty  } \frac{d y}{y} Sech(y /2)^4 ( - 2 y + y
Cosh(y ) +  Sinh(y ) ). \ee
Note that this integral is convergent.
%
%
Numeric integration in Eq.\Ref{I3a} gives
\be \label{I3num} I_3 = - A_0 \pi \beta ~(0.3348). \ee
In such a way all the other integrations in Eqs.\Ref{M1}, \Ref{M2}
can be carried out.

Performing analogous calculations for other terms we obtain the
  expression for scattering amplitude in high
temperature approximation. A not complicated problem is to find
next-to-leading corrections having the order $(m \beta)^l, l = 1,
2, ...$ As a result, the explicit high temperature limits  for the
scattering amplitude can be calculated in terms of elementary
functions.
\section{Scattering of  photons on the induced potential
$\bar{\phi}^3$, $\bar{\phi}^8$ }

Already, in Eqs.\Ref{solution}, \Ref{Fphi3} the expressions for
the  potential $\bar{Q}^3_4 = \bar{\phi}^3$ generated in the
  stationary plasma plate were calculated.
%
%
%
%
%
%
Since $ \bar{\phi}^3$ is the classical field, we could consider
scattering of classical  electromagnetic waves on the plasma.
However here we calculate the scattering of photons on the
potential \Ref{Fphi3}.

Let us denote the momentum of ingoing photon as $k^1_\mu$ and the
one for outgoing photon as  $k^3_\la.$ The matrix element of the
process reads
\be \label{M} M = (2 \pi)^4 \delta(k^1 + k^2  - k^3)
~~\frac{e^{\sigma_1}_\mu}{\sqrt{2 \omega_1}} ~ \bar{\phi}^3 ~
\Gamma^4_{\mu\la}~\frac{e^{\sigma_3}_\la }{\sqrt{2 \omega_3}}. \ee
Here, ${e^{\sigma_1}}_\mu, {e^{\sigma_2}}_\la $ are polarization
amplitudes and
 of photons, $\omega_1, \omega_3$ - corresponding energies. $ \Gamma^4_{\mu\la}(k^1,
 k^3)$ is effective vertex calculated in previous section.

 We assume the not polarized beams,
\be \sum_{\sigma_1} e^{\sigma_1}_\mu  e^{\sigma_1}_{\mu'}=
\delta_{\mu \mu'} ~ \sum_{\sigma_3} e^{\sigma_3}_\la
e^{\sigma_3}_{\la'}= \delta_{\la \la'}, \ee ~
 and obtain for the probability

\be \label{P} P = M M^+ = ( \bar{\phi}^3(k) )^2~ \Gamma^4_{\mu\la}
\Gamma^4_{\mu\la} \frac{C}{4 \omega_1 \omega_3}  ~\delta(k^1 + k^2
- k^3), \ee
where C is a not relevant now number.   In this expression,
accounting for the momentum conservation, $\omega_3 =[
(\omega^1_x)^2 + (\omega^1_y )^2 + (\omega^1_z  +
k^2_z)^2]^{1/2}$. The  value of $k^2_z$ is free parameter of the
problem. It measures at which point of the z-plane the actual
scattering happens. Since this is not known, we have to sum up the
probability in $k^2_z$, that is in $l$. In this expression all the
parameters and functions are known. So, the scattering on the
 induced color potentials can be calculated. Analogous
process has to happen due to the classical field $
\bar{\phi}^8(k)$. This kind of scattering drastically differs from
that for the plasma consisting of free particles.

Other related process is the conversion of classical gluon fields
$ \bar{\phi}^3(k),  \bar{\phi}^8(k)$ in two photons coming out
from the QGP due to color parity violation and  the effective
vertex $ \Gamma^\nu_{\mu\la}(k^1, k^3)$. In the rest frame of the
plasma, two photons  moving in opposite directions and having
specific energies, which  correspond  to the energy levels $E_l$
Eq.\Ref{emode}, have to be observed.

Both these processes  could serve as the signals of the DPT.

\section{Appendix} Below we present Feynman's rules used. They
have been derived from FeynArts package by using the rotation to
the Euclid space-time for momentum and coordinates: $p_0 \to i
p_4, x_0 \to - i x_4$.

The fermion propagator in Minkowsky space-time reads
\be \label{G} G ( x) = \int \frac{d^4 p}{(2\pi)^4} \frac{ i
(\hat{p} + m )~ e^{- i px}}{p^2 - m^2 + i 0}. \ee
Here  $\hat{p} = \gamma_\mu  p^\mu = \gamma_0 p_0 -
\vec{\gamma}\vec{p}, p^2 = p_0^2 - \vec{p}^2 $.

The transformation of the $\gamma$ matrixes is established
analogously  to the coordinates: $ \gamma_0 \to - i \gamma_4$.
That results in the anti-commutation relation
\be \label{rgamma} \gamma^E_\mu  \gamma^E_\nu +   \gamma^E_\nu
\gamma^E_\mu = - 2 \delta_{\mu\nu}. \ee
That is $g_{\mu\nu} = -  \delta_{\mu\nu}$. Fulfilling this
transformation and the shift $ \vec{p} \to  - \vec{p}$, we obtain
$\gamma_\mu  p^\mu \to \gamma_\mu^E  p_\mu^E =   \gamma_4 p_4  +
\vec{\gamma}\vec{p}$.  Note that in the Euclid space-time we have:
\be \gamma_\mu^+ = - \gamma_\mu .\ee
 For the transformed propagator we obtain
\be \label{Ge} G_E ( x) = \int \frac{d^4 p_E}{(2\pi)^4} \frac{
(\hat{p}_E + m )~ e^{- i p_Ex_E}}{p_E^2 +  m^2 }. \ee

Now we consider interactions. In the  Minkowsky space-time the
action is
\be \label{actionM} A = \int d^4 x  [i \bar{\psi} \gamma^\mu
(\partial_\mu - i e A_\mu - i g \frac{\lambda^a}{2} Q^a_\mu ) \psi
- m \bar{\psi} \psi ], \ee
where one quark flavor is considered and $A_\mu,  Q^a_\mu$ are
electromagnetic  potential and gluon field potential, e, q are
electric and strong charges of quark, $\lambda^a$, a = 1, ..., 8,
are the Gell-Mann  matrices, color indexes are  assumed.

Performing  the transformation of coordinates, $\gamma_0 $ matrix
and defining the component $A_4 = i A_0$, we obtain
\be \label{vertAe} A_{em} = - \int d^4 x_E~[- e \bar{\psi_E}
\gamma^\mu_E \psi_E A_\mu^E]. \ee

The expression for the quark-gluon interaction is
\be \label{vertQe}  A_{str.} = - \int d^4 x_E~[- g \bar{\psi_E}
\frac{\lambda^a}{2} \gamma^\mu_E \psi_E ( Q^a_\mu)^E]. \ee

\end{document}